\documentclass[journal,10pt,twocolumn,twoside]{IEEEtran}
\usepackage{amsmath,amsfonts,amssymb,amsbsy,bm,paralist,theorem,color}
\usepackage{graphicx,algorithmic,algorithm,relsize}
\usepackage{multicol}
\usepackage{multirow}
\usepackage{subfigure}
\usepackage{caption}
\usepackage{cite}

\graphicspath{{fig/}}

\definecolor{orange}{RGB}{255,107,0}
\definecolor{green}{RGB}{0,160,20}

\begin{document}
\title{A Reinforcement Learning Approach for an IRS-assisted NOMA Network }
\author{Ximing Xie, Shiyu Jiao and Zhiguo Ding,~\IEEEmembership{Fellow,~IEEE}
\thanks{Ximing Xie, Shiyu Jiao and Zhiguo Ding are with School of Electrical and Electronic Engineering, The University of Manchester, M13 9PL, U.K. (e-mail: ximing.xie@manchester.ac.uk, shiyu.jiao@manchester.ac.uk and zhiguo.ding@manchester.ac.uk).}
}\maketitle

\begin{abstract}
This letter investigates a sum rate maximization problem in an intelligent reflective surface (IRS) assisted non-orthogonal multiple access (NOMA) downlink network. Specifically, the sum rate of all the users is maximized by jointly optimizing the beams at the base station and the phase shift at the IRS. The deep reinforcement learning (DRL), which has achieved massive successes, is applied to solve this sum rate maximization problem. In particular, an algorithm based on the deep deterministic policy gradient (DDPG) is proposed. Both the random channel case and the fixed channel case are studied in this letter. The simulation result illustrates that the DDPG based algorithm has the competitive performance on both cases.
\end{abstract}
\vspace{-0.8em}
\section{Introduction}
The ultra-massive machine type communication (umMTC) is a key scenario of the next generation mobile communication \cite{you2021towards}. An umMTC network always consists of massive communication devices, e.g. mobiles and sensors. Each device will communicate with other devices and cause massive traffic. Since the spectrum resource is extremely restricted, it is a challenge to support such heavy traffic. The non-orthogonal multiple access (NOMA) is being considered as a potential candidate of 6G communication system \cite{ding2015application}. The main feature of NOMA is that it allows multiple devices share the same spectrum resource to communicate simultaneously, which greatly improves the spectrum efficiency. In particular, to satisfy the individual quality of service (QoS) requirement, each user or device in a NOMA network always adopts successive interference cancellation (SIC) to improve the signal to interference and noise ratio (SINR) and reception reliability \cite{saito2015performance}. \par

Recently, the intelligent reflective surface (IRS) is proposed as a potential auxiliary device to improve the channel quality and help users in the area with heavy blockage to receive signals\cite{8811733}. An IRS consists of many passive reflecting elements and a smart controller which can adjust the phase shift of each reflecting element. The IRS can adapt the channel between the transmitter and the receiver to increase the channel gain. Moreover, the IRS performs as a mirror to redirect the signal and enlarge the signal prorogation range. \par

Since the artificial intelligence has achieved great success, especially deep learning (DL), in wireless communication field  \cite{huang2020reconfigurable}, many works have studied the application of  DL algorithms in a wireless communication network \cite{gao2020unsupervised}. Deep reinforcement learning (DRL) as a type of DL has been attracting more and more attentions since there is no training data requirement \cite{arulkumaran2017deep}. Unlike the supervised learning and unsupervised learning, for which the training data is crucial, DRL adopts an agent to 
continuously interact with the environment and fetch feedbacks and utilizes the feedback to train the agent. \par

In this letter, a sum rate maximization problem is investigated and an algorithm based on DDPG is proposed. Due to the non-convexity of the formulated problem, it is very challenging to find the optimal solution via conventional convex optimization. The DDPG based algorithm is proposed as an alternative method to efficiently solve this problem. This letter also investigates two types of channels, which are the fixed channel and the time-varying channel. The simulation results demonstrate that the DDPG algorithm can adapt both types of channels and has the superior performance on both cases.
\section{System Model and Problem Formulation}

\begin{figure}[tp]
	\centering
	\includegraphics[width=0.6\linewidth]{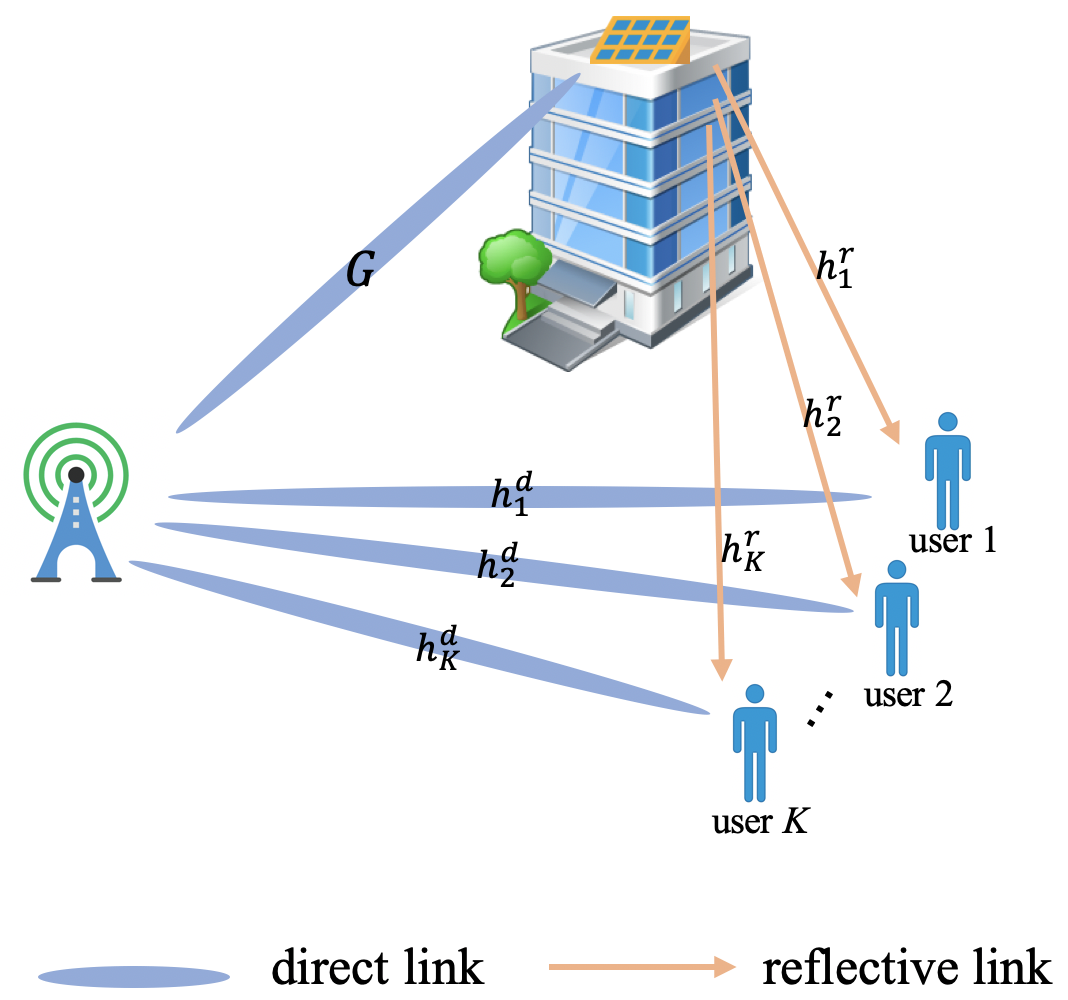}\\
		\captionsetup{font={small}}
	\caption{IRS NOMA sytem model.} \label{system model}
	\vspace{-0.8cm}
\end{figure}

We consider a NOMA-MISO downlink system comprising a BS and a IRS as shown in Fig. \ref{system model}. The BS is equipped with $M$ antennas and the IRS has $N$ reflective elements and a smart controller. The BS communicates with $K$ ($M \leq K$) single antenna users. The user set is defined as $\mathcal{K} = \{1,2,...,K\}$. It is assumed that the IRS is deployed on the surface of a building, thus the distance between the BS and the IRS is fixed. As shown in Fig. \ref{system model}, each user will receive the reflected signal from the IRS and the direct signal from the BS. The BS will generate an unique beam for each user and superimpose each user's signal. The superimposed 
signal sent by the BS can be expressed as follows:
\begin{equation}
y = \sum\limits_{k=1}^{K}\mathbf{w}_kx_k, \label{sent signal}
\end{equation}
where $y$ denotes the superimposed signal sent to all the users, $\mathbf{w}_k \in \mathbb{C}^M$ denotes the beamforming vector for the $k$-th user and $x_k$ denotes the signal symbol of $k$-th user. We assume the power of signal symbol is unity which means $\mathbb{E}(x_k^2) = 1, \forall k$. The received signal of $k$-th user can be expressed as follows:
\begin{equation}
y_k =  \underbrace{\mathbf{h}_k^{dH}y}_{\text{\rm direct link}} + \ \underbrace{\mathbf{h}_k^{rH}\mathbf{\Phi}\mathbf{G}y}_{\text{\rm reflective link}} + w_k, k \in \mathcal{K},\label{received signal}
\end{equation}
where $\mathbf{\Phi} \triangleq {\rm diag}[\phi_1, \phi_2, ..., \phi_N]$ denotes the phase shift matrix of the IRS and $w_k$ is the zero mean additive white Gaussian noise (AWGN) with variance $\sigma_n^2$. $\mathbf{G} \in \mathbb{C}^{N \times M}$ denotes the channel matrix between the BS and the IRS.  $\mathbf{h}_k^d \in \mathbb{C}^{M \times 1}$ and $\mathbf{h}_k^r \in \mathbb{C}^{N \times 1}$ denote the direct link and the reflective link of the $k$-th user, respectively.\par

Note that $\mathbf{\Phi}$ is a diagonal matrix and each element on the 
main diagonal describes the state of a reflecting element on the IRS. The element on the main diagonal is expressed as $\phi_n = \beta_n e^{j\theta_n}$, where $\beta_n \in [0,1]$ is the reflecting coefficient describing the signal energy loss at the IRS and $\theta_n \in [0,2\pi]$ is the phase shift introduced by the reflecting element. It is assumed that the IRS performs ideal reflection that the energy loss caused by reflection is ignored. Thus, the reflected signal will not suffer energy loss at the IRS, which means $\beta_n = 1, \forall n$ or $|\phi_n| =1, \forall n$.

For the simple notation, $\hat{\mathbf{h}}_k = \mathbf{h}_k^{dH} + \mathbf{h}_k^{rH}\mathbf{\Phi}\mathbf{G}$ is utilized to represent the composite channel of the $k$-th user. To successfully apply SIC, the first step is to decide the decoding order.  In this letter, channel quality is considered as the factor to decide the decoding order. If the composite channels are sorted as $|\hat{\mathbf{h}}_1|^2 \leq |\hat{\mathbf{h}}_2|^2 \leq ... \leq |\hat{\mathbf{h}}_{K-1}|^2 \leq |\hat{\mathbf{h}}_K|^2$, then the decoding order is defined as $\bar{\epsilon}=(1,2,...,K)$. Then, the interference set of the $k$-th user is defined as $\kappa_k$, which contains the index of all the users causing interference to the $k$-th user. For example, considering the decoding order $\bar{\epsilon}$, the $k$-th user will decode the first $k-1$ users' signals and remove them from the received signal. Thus, only users after the $k$-th user will cause interference to the $k$-th user when the $k$-th user decodes its own data. Therefore, the interference set of the $k$-th user based on $\bar{\epsilon}$ is $\{k+1,...,K\}$. It is worth to point out that the interference set of each user is various according to different decoding orders. We use $\gamma_{ij}$ to represent the user $i$'s SINR observed at user $j$, which can be expressed as follows:
\begin{equation}
\gamma_{ij} = \frac{|\hat{\mathbf{h}}_j \mathbf{w}_i|^2}{\sum\limits_{k \in \kappa_i}|\hat{\mathbf{h}}_j\mathbf{w}_k|^2 + \sigma_n^2},\label{sinrij}
\end{equation}
and the user $i$'s data rate observed at user $j$ is given by:
\begin{equation}
R_{ij} = \log_2(1+\gamma_{ij}).\label{rate_ij}
\end{equation}
The user $i$'s SINR when it decodes its own data can be expressed as follows:
\begin{equation}
\gamma_i = \frac{|\hat{\mathbf{h}}_i \mathbf{w}_i|^2}{\sum\limits_{k \in \kappa_i}|\hat{\mathbf{h}}_i\mathbf{w}_k|^2 + \sigma_n^2},\label{sinr_i}
\end{equation}
and the user $i$'s data rate is given by:
\begin{equation}
R_{i} = \log_2(1+\gamma_{i}).\label{rate_i}
\end{equation}\par

The aim of this letter is to maximize the sum rate of all the users. The optimization problem can be formulated as follows:
\begin{subequations}\label{Prob0} 
\begin{align}
&\max_{\{\mathbf{w},\mathbf{\Phi}\}}\quad\sum\limits_{i=1}^K R_i \label{P00}\\
		&~\mathrm{s.t.}\qquad \min(R_{ij}, R_i) \geq R_t, \forall j \in \kappa_i, i \in \mathcal{K} \label{P01}\\
		&\qquad\qquad \sum\limits_{i=1}^{K} |\mathbf{w}_i|^2\leq P_t \label{P02}\\
		&\qquad\qquad ||\mathbf{\Phi}_{n,n}|| = 1, \forall n \label{P03}\\
		&\qquad\qquad 0 \leq \theta_n \leq 2\pi, \forall n. \label{P04}
\end{align}
\end{subequations}
The constraint \eqref{P01} guarantees that SIC can proceed smoothly, where $R_t$ is the target data rate. It is assumed that all the users have the same target rate. The constraint \eqref{P02} is related to power control, which indicates that the total transmit power cannot exceed the system maximum power $P_t$. The constraints \eqref{P03} and \eqref{P04} are introduced by the IRS. $\mathbf{\Phi}_{n,n}$ denotes the $n$-th element on the main diagonal of the phase shift matrix $\mathbf{\Phi}$ and \eqref{P03} indicates that the IRS is an ideal reflective mirror and will not introduce any noise. \par

The formulated problem is non-convex due to the non-convex objective function \eqref{P00} and the non-convex constraint \eqref{P01}. Multiple optimization valuables are coupled together in \eqref{P00} and \eqref{P01}. It is difficult to find the global optimal solution through conventional convex optimization. There are some existing works using the alternating algorithm  \cite{zuo2020resource} and  zero forcing (ZF) \cite{ding2020simple} to find a sub-optimal solution with the assumption that the channel is fixed. In the next section, a DDPG based algorithm is proposed to solve the aforementioned problem efficiently. Even the channel is time varying, the DDPG based algorithm can still achieve competitive performance.

\section{DDPG-based Joint Optimization of Phase Shift and Beamforming}

\subsection{Basic knowledge of DDPG }
The reinforcement learning (RL) is an area of machine learning that handles with sequential decision-making \cite{franccois2018introduction}. There are two critical parts for a RL system, which are the agent and the environment. The key idea of RL is to train an agent to generate good actions based on the environment. A few factors fully characterize the RL processing. Policy $\pi$ reflects the probability of an action chosen by an agent. State $s$ is an observation from the environment. Action $a$ is the decision made by the agent. Reward $r$ is the feedback of an action.
There are two kinds of RL algorithms in the RL family, one is value-based RL and another is policy gradient RL. \par

The value-based RL algorithm aims to build a value function and then decide a policy by minimizing or maximizing this function. The deep Q learning (DQN) is a typical value-based RL. In DQN, the loss function is defined as follows:
\begin{equation}
L(\theta)=\mathbb{E}\left[\left(r_t +\xi \max _{a^{\prime}} Q^{\prime}\left(s_{t+1}, a^{\prime} | \theta^{\prime}\right)-Q(s_t, a_t | \theta)\right)^{2}\right], \label{loss function}
\end{equation}
where $Q'$ is the target network and $Q$ is the training network. It is worth to point out the parameters of the target network $\theta^{\prime}$ are fixed during the training. DQL is designed to address discrete action problems.\par

The policy gradient (PG) RL algorithm aims to optimize a performance objective by finding a good policy \cite{franccois2018introduction}. PG algorithms are also designed for the continuous action space. There is one classic PG algorithm named stochastic policy gradient. The parameter updating rule can be expressed as follows:

\begin{equation}
\theta_{t+1} = \theta_t + l\mathbb{E}[\nabla_aQ(s_t,a_t)| \; \nabla_\theta\mu_\theta(s_t)], \label{stochastic pg}
\end{equation}
where $l$ is the learning rate, $Q^{{\pi_\theta}}$ is the $Q$ function to evaluate the current policy and $\mu_\theta$ is the current policy. \par

The idea of DDPG combines DQN and PG together which can handle with the problem with a continues action space and also has a better convergence performance. A DDPG framework is constructed by two training networks and two target networks. The training actor network performs as a policy to generate actions. The training critic network estimates the Q function to evaluate the action. The target actor network generates the estimated action $a'$ in \eqref{loss function}. The target critic network generates the estimated Q value $Q'$ in \eqref{loss function}. \eqref{loss function} and \eqref{stochastic pg} are utilized to train the training critic network and the training actor network, respectively. Note that the target network shares the same structure with its associated training network. Therefore, the target networks are updated through soft update, which is given by
\begin{equation}
\theta^{(target)} = \tau\theta^{(train)} + (1-\tau)\theta^{(target)}, \label{soft update}
\end{equation}
where $\tau$ is the updating rate .

\subsection{Proposed DDPG framework}

The corresponding elements are defined as follows:
\begin{itemize}
\item action $a^{(t)} = \left[\mathbf{w}_1^{(t)},...,\mathbf{w}_K^{(t)} , \mathbf{\Phi}^{(t)}\right]$
\item state $$s^{(t)}=\left[\gamma_1^{(t-1)},..., \gamma_K^{(t-1)}, a^{(t-1)},\\
|\mathbf{w}_1^{(t-1)}|^2,...,|\mathbf{w}_K^{(t-1)}|^2\right]$$
\item reward $r^{(t)} = \sum\limits_{i=1}^K R_i^{(t)}$
\end{itemize}
Note that the state space contains the square norm of each beamforming vector. It is reasonable because the neural network should take energy consumption into consideration due to the power control constraint \eqref{P02}. It is worth to point out that the neural network can only take real number, therefore, the real part and the image part of a complex number should be separately input into neural networks.\par

In this letter, both the actor network and the critic network are fully connected neural networks, comprised of input layer, hidden layer, batch normalization layer and output layer. The size of the actor network's input layer is determined by the dimension of the state tuple. The critic network has two input layers for the action and the state specifically. The outputs of these two input layers will be horizontally stacked together as the input of the next hidden layer. The size of the hidden layer is related to the number of antennas at the BS, the number of reflecting elements at the IRS and the number of users. In this letter, we adopt 300 neurons in every hidden layer. The batch normalization layer is utilized between two hidden layers to contribute to faster convergence and shorter training time. The active functions utilized in the proposed DDPG are $tanh$ and $relu$, which make back propagation and gradient decent easier. Adam optimizer is utilized for both two networks with the learning rates 0.001 for the actor network and 0.002 for the critic network, respectively. \par

The experience replay is adopted in this DDPG framework to reduce the correlation of different training
samples. A replay buffer $\mathcal{M}$ with the capacity $\mathcal{C}$ is implemented at the beginning of training. The training sample of each step, which is constructed by $\{a^{(t)}, r^{(t)}, s^{(t)}, s^{(t+1)}\}$, is stored into the replay buffer. The training progress will only start until the replay buffer is full. If the replay buffer is full, the newest training sample will replace the earliest one. In each training step, a mini-batch training samples are randomly selected from the replay buffer as the training data, which guarantees that the DDPG model has a good view of every training step.

A constraint fulfil layer is implemented as the output layer of the actor network, which guarantees that the action output from the actor network must satisfy all the constraints in \eqref{Prob0}. The next subsection introduces how this layer handles with all constraints.
\subsection{Constraint handling}
In the conventional optimization algorithms, for instance the SDR algorithm and the alternating algorithm, we usually convert a non-convex problem to a convex problem and then solve it via convex optimization solver, like CVX in Matlab. CVX can directly handle with convex constraints. However, this is not acceptable in deep neural networks. We manipulate the output layer of the actor network to make the action satisfy all the constraints in \eqref{Prob0}. \par
It is known that power and the data rate are positively related from Shannon's Theory. Higher data rate requires more energy consumption. Note that the maximum transmit power at the BS is $P_t$ from \eqref{P02}. Therefore, only letting the total transmit power at the BS as the maximum power allowed by the system can obtain the maximum sum rate. Hence, the constraint \eqref{P02} becomes an equality, which is given by:
\begin{equation}
\sum\limits_{i=1}^{K} |\mathbf{w}_i|^2 = P_t. \label{P03'}
\end{equation}
$\mathbf{w}_i^{(l)}$ denotes the beamforming vector for $i$-th user obtained by the actor network in the $l$-th training step. Since there is no constraint handling inside the neural network, thus  $\mathbf{w}_i^{(l)}, \forall i$ may not satisfy the constraint \eqref{P03'}. Therefore, normalization is utilized to guarantee that constraint \eqref{P03'} is not violated. In the $l$-th step, we first calculate the total transmit power directly obtained by the actor network, which is
\begin{equation}
P^{(l)} = \sum\limits_{i=1}^{K} |\mathbf{w}_i^{(l)}|^2.
\end{equation}
Then, we reassign power to user $i$'s beam based on the following rule:
\begin{equation}
\mathbf{w}_{i}^{*(l)} = \mathbf{w}_i^{{(l)}}\sqrt{\frac{P_t}{P^{(l)}}}. \label{beam norm}
\end{equation}
The key idea of \eqref{beam norm} is reassigning the power based on the ratio $\frac{|\mathbf{w}_i^{(l)}|^2}{P^{(l)}}$ and setting the total transmit power as $P_t$. Meanwhile, the new beamforming vector $\mathbf{w}_i^{*(l)}$  has the same direction with $\mathbf{w}_i^{(l)}$. After the operation of \eqref{beam norm}, all the new beamforming vector satisfy
\begin{equation}
\sum\limits_{i=1}^{K} |\mathbf{w}_i^{*(l)}|^2 = P_t. 
\end{equation}
\par

As for the constraint \eqref{P03}, normalization is used again to handle it. $\mathbf{\Theta}^{(l)} = \{\phi_1^{(l)},\phi_2^{(l)},...,\phi_N^{(l)}\}$ containing all the elements on the main diagonal of the phase shift matrix denotes the optimized result directly obtained by the actor network in the $l$-th step. To satisfy the constraint \eqref{P03}, each element can be normalized by the following
\begin{equation}
\phi_i^{*(l)} = \frac{\phi_i^{(l)}}{|\phi_i^{(l)}|}, \quad i = \{1,2,...,N\}. \label{theta norm}
\end{equation}
It is known that the function $f(\theta) = e^{j\theta}$ is a periodic function with a period of $2\pi$. Therefore, the phase can be always mapped within $[0, 2\pi]$. \par
In the conventional optimization, it is difficult to handle with the constraint \eqref{P01} since it is non-convex and contains two optimization valuables coupled together. Conventionally, SDR is utilized to approximate it and then the alternating algorithm is applied to find a suboptimal solution. In this DPL-based algorithm,  we have $\mathbf{w}_i^{*{l}}, \forall i$ and $\Theta^{*(l)}$ from \eqref{beam norm} and \eqref{theta norm}, respectively. It is known that one user has an original data rate and several observed date rates in a NOMA system. Once a specific decoding order $\epsilon^{'}$, $\mathbf{w}_i^{*{l}}, \forall i$ and $\Theta^{*(l)}$ are decided, the original data rate and the observed data rates can be calculated via \eqref{rate_i} and \eqref{rate_ij} for each user, respectively. Then, we check if each user's original data rate and observed data rates satisfy the constraint \eqref{P01}. If the constraint \eqref{P01} is violated, it means the actor networks outputs an invalid action. In order to reduce the invalid output as much as possible, a punishment mechanism is adopted on the reward. If the constraint \eqref{P01} is violated, the reward is set as
\begin{equation}
r^{(l)} = \sum\limits_{i \in \mathcal{K}}\sum\limits_{j \in \kappa_i}\min(\min(R_{ij}^{(l)}, R_i^{(l)}) -R_t,0)\label{neg reward}
\end{equation}
instead of the sum rate. Note that the reword is negative when constraint \eqref{P01} cannot be satisfied in the $l$-th training step. Meanwhile, if the action seriously violates the \eqref{P01}, the reword will be assigned a smaller negative value, which is a more severe punishment to the neural network. Hence, the neural network will adjust the output in the following training steps to avoid such invalid action as much as possible.

\subsection{Algorithm}
The detail of the algorithm is shown in  {\bf Algorithm} \ref{Alg}. At the beginning of the algorithm, four networks and the replay buffer are initialized. At the beginning of each training episode, CSI, beamforming vectors and phase shift matrix are initialized. We simply adopt an identity matrix to initialize all the beamforming vectors. Note that CSI in each episode is various, which indicates that this DDPG model is compatible with the varying channel scenario. In each training step, a random process $\mathcal{N}_t$ is adopted to enlarge the action exploration. $\mathcal{N}_t$ in {\bf Algorithm} \ref{Alg} is a complex Gaussian noise vector with the same size of the action $a$. It is worth to point out that the composite channel of each user is changing due to various phase shifts in one training episode. As mentioned before, the decoding order is decided by the composite channel gain, therefore, the new decoding order needs to be calculated based on the current phase shift, which is shown in step 11.
\begin{algorithm}[tp]
	\caption{DDPG-based Joint Beamforming and Phase Shift Optimization}\label{Alg}
	\begin{algorithmic}[1] 
		\STATE {{\bf Initialization:} Generate two actor networks and two critic networks. Make the training network and the target network have the identical parameters, $\theta_a^{(train)} = \theta_a^{(target)}$ and $\theta_c^{(train)} = \theta_c^{(target)}$. \\
		Initialize the experience replay buffer $\mathcal{M}$ with the storage capacity $C$ and the mini-batch size $N_b$. }
		\STATE {{\bf Output:} Optimal beamforming vector $\mathbf{w}_k, k \in \mathcal{K}$ and phase shift matrix $\mathbf{\Phi}$.}
		\FOR { episode $i = 1,2,...,I$ }
		\STATE {Randomly generate CSI $\mathbf{G}^{(i)}$, $\mathbf{h}_k^{d(i)}, \mathbf{h}_k^{r(i)}, k \in \mathcal{K}$ and the phase shift matrix $\Phi^{(i)}$. The beamforming vectors are initialized by a $M \times K$ identity complex matrix. }
		\STATE {Calculate the decoding order $\epsilon_0$ according the the composite channel power.}
		\STATE {Obtain the initial state $s_1$}
			\FOR { step $t = 1,2,...,T$ }
			\STATE {Initialize a random process $\mathcal{N}_t$. }
			\STATE {Choose an action from the actor training network $a_t = \mu^{(train)}(s_t|\theta_a^{(train)})+\mathcal{N}_t$.}
			\STATE {Normalization the beamforming vectors and the phase shift matix by \eqref{beam norm} and \eqref{theta norm}. }
			\STATE {Calculate the new decoding order $\epsilon_t$.}
			\STATE {If constraint \eqref{P01} is not violated, the reward $r_t$ is set as the sum of all users' original data rate. Otherwise, $r_t$ is given by \eqref{neg reward}. Obtain the new state $s_{t+1}$.}
			\STATE {Store $\{s_t,a_t,r_t,s_{t+1}\}$ to the buffer $\mathcal{M}$.}
			\STATE {Sample a minibatch with the batch size $N_b$ from $\mathcal{M}$ to train networks.}
			\STATE {Given reword discount factor $\xi$, set the target Q value based on \eqref{loss function}.}
			\STATE {Update $Q^{(train)}(s,a|\theta_c^{(target)})$ by minimizing the loss function.}
			\STATE {Update $\mu^{(train)}(s|\theta_a^{(train)})$ by the policy gradient. }
			\STATE {Update two target networks by using soft update.}
			\STATE {$s_t = s_{t+1}$.}
			\ENDFOR
		\ENDFOR
		\STATE {{\bf Output} \{$\mathbf{w}_k^{*(j)}$, $\alpha_k^{*(j)}$, $\mathbf{e}_k^{*(j)}$\}, $\forall k$.}
	\end{algorithmic}\label{Al1}
	\vspace{-0.1cm}
\end{algorithm}

\section{Numerical Results}
This section demonstrates the performance of the proposed algorithm. It is assumed that the channels between the BS and all the users are Rayleigh channel, which indicates that the line-of-sight (LoS) signal is blocked and the users mainly receive signals through the IRS. This channel is model as follows:
\begin{equation}
\mathbf{h}_{k}^d =  \widetilde{\mathbf{h}}_{k}^d / \sqrt{d_{d,k}^\alpha}, k \in \mathcal{K},
\end{equation}
where $\widetilde{\mathbf{h}}_{d,k} \in \mathbb{C}^{M \times 1}$ contains $M$ independent and identical
elements following complex $\mathcal{CN}(0,1)$ distribution. $d_{d,k}$ denotes the distance between the BS and the $k$-th user, which is generated randomly within the range (45,50). $\alpha$ denotes the path loss exponent, which is set as 2. The channel between the BS and the IRS and the channels between the IRS and all the users are assumed to be Rician channel, which can be modelled as follows:
\begin{align}
&\mathbf{G} = \left(\sqrt{\frac{\upsilon}{1+\upsilon}}\mathbf{G}^{\rm LoS} +  \sqrt{\frac{1}{1+\upsilon}}\mathbf{G}^{\rm nLoS}\right)/\sqrt{d^\alpha}, \\
&\mathbf{h}_{k}^r = \left(\sqrt{\frac{\upsilon}{1+\upsilon}}\mathbf{h}_{k}^{r\rm LoS} +  \sqrt{\frac{1}{1+\upsilon}}\mathbf{h}_{k}^{r\rm nLoS}\right)/\sqrt{d_{r,k}^\alpha}, k \in \mathcal{K}, 
\end{align}
where $\upsilon$ denotes the Rician factor,  $\mathbf{G}^{\rm LoS}$ and $\mathbf{h}_{k}^{r\rm LoS}$ are the LoS component, $\mathbf{G}^{\rm nLoS}$ and $\mathbf{h}_{k}^{r\rm nLoS}$ are the non-LoS component and $d$ and $d_{r,k}$ denote the distance between the BS and the IRS and the distance between the IRS and the $k$-th user. In this letter, the Rician factor is set as 1 and the LoS components $\mathbf{G}^{\rm LoS}$ and $\mathbf{h}_{k}^{r\rm LoS}$ are assumed to be 1. The non-LoS components  $\mathbf{G}^{\rm nLoS}$ and $\mathbf{h}_{k}^{r\rm nLoS}$ follow the Rayleigh fading. $d$ is set as 50 and $d_{r,k}$ is randomly generated within the range (5,10). \par

\begin{figure}[tp]
\subfigure[Varying channel]{
\begin{minipage}[t]{0.5\linewidth}
\centering
\includegraphics[width=2in]{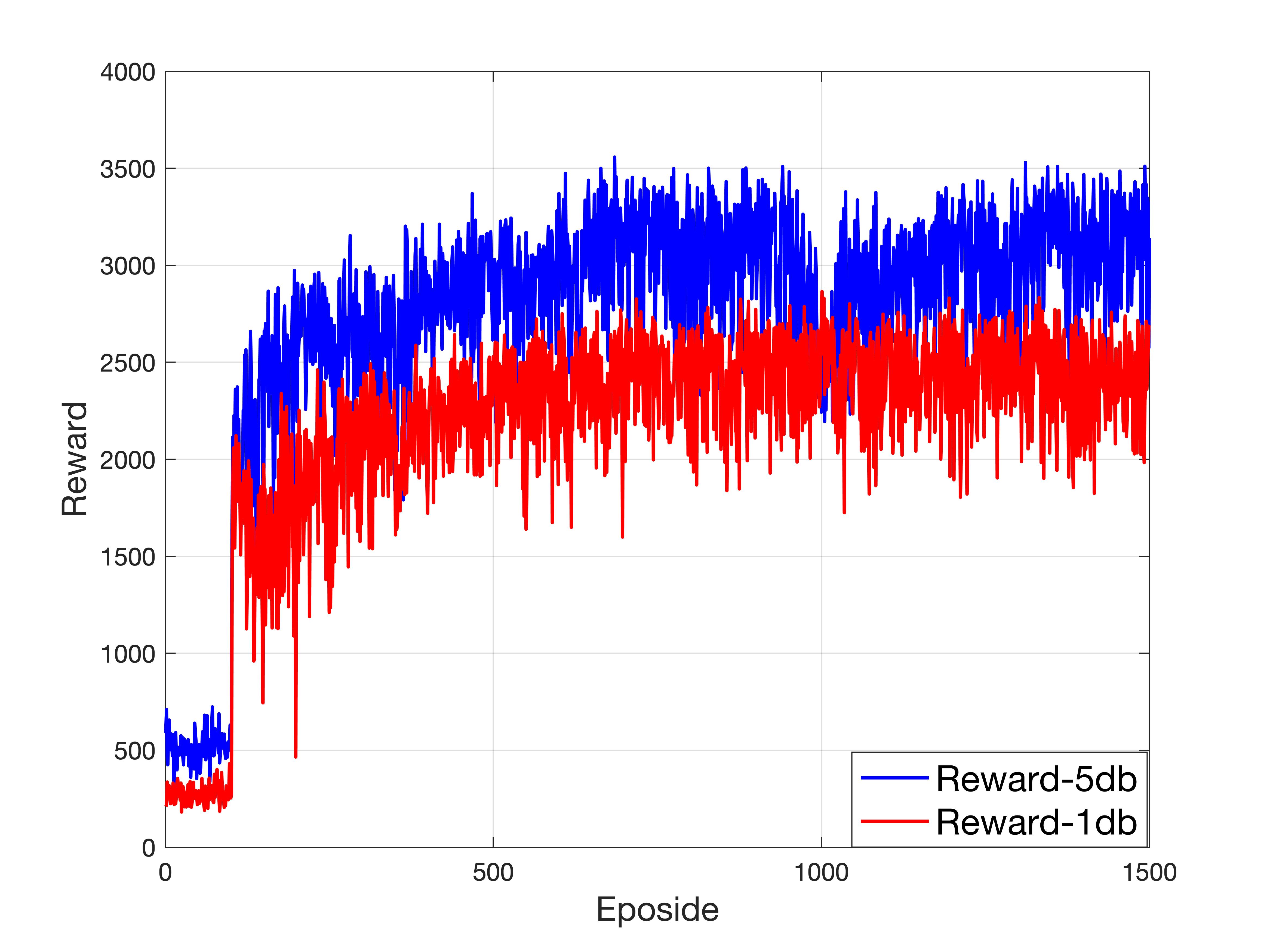}
\end{minipage}%
}%
\subfigure[Fixed channel]{
\begin{minipage}[t]{0.5\linewidth}
\centering
\includegraphics[width=2in]{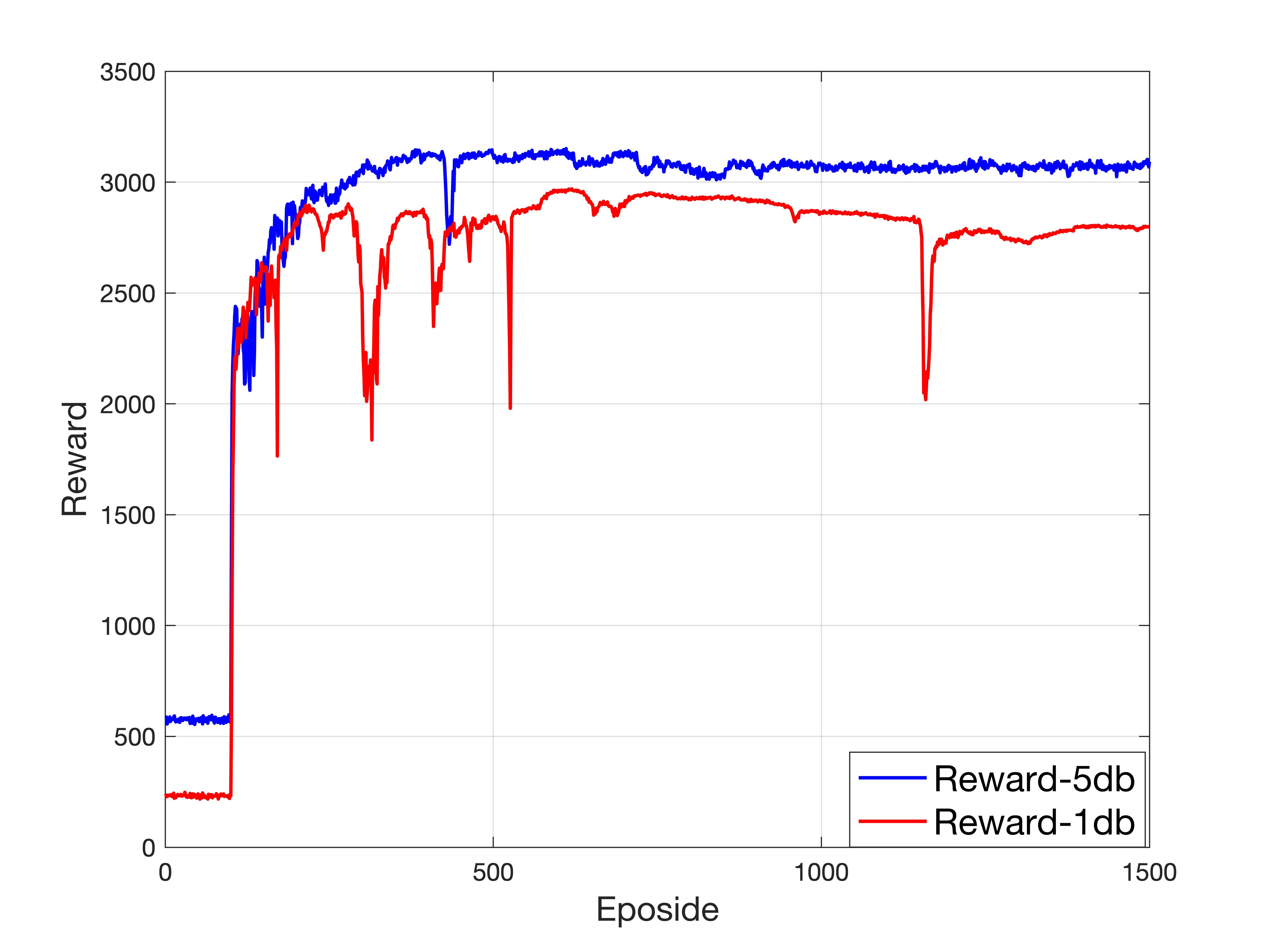}
\end{minipage}%
}%
\captionsetup{font={small}}
\caption{The accumulative reward } \label{reward}
\vspace{-0.6cm}
\end{figure}

Fig. \ref{reward} shows how the accumulative reward changes during training. In this simulation, the number of users is 4, the number of IRS elements is 32 and the number of anteenas at the BS is 4. The noise power is $\sigma^2 = -10$ dBm. The subfigure (a) illustrates the scenario that the channel is varying. It shows the accumulative reward converges with the training proceeds, which indicates that the proposed DDPG algorithm has a good adaptation to time-varying channels. The subfigure (b) illustrates the scenario that the channel is fixed. Compared with the varying channel case, the reward is more stable. Note that the training samples are randomly selected from the replay buffer at the beginning of each episode. If the selected samples are not good experience, i.e., action violates the constraint, the reward will drop suddenly at this episode. \par

\begin{figure}[tp]
\subfigure[Transmit power]{
\begin{minipage}[t]{0.5\linewidth}
\centering
\includegraphics[width=2in]{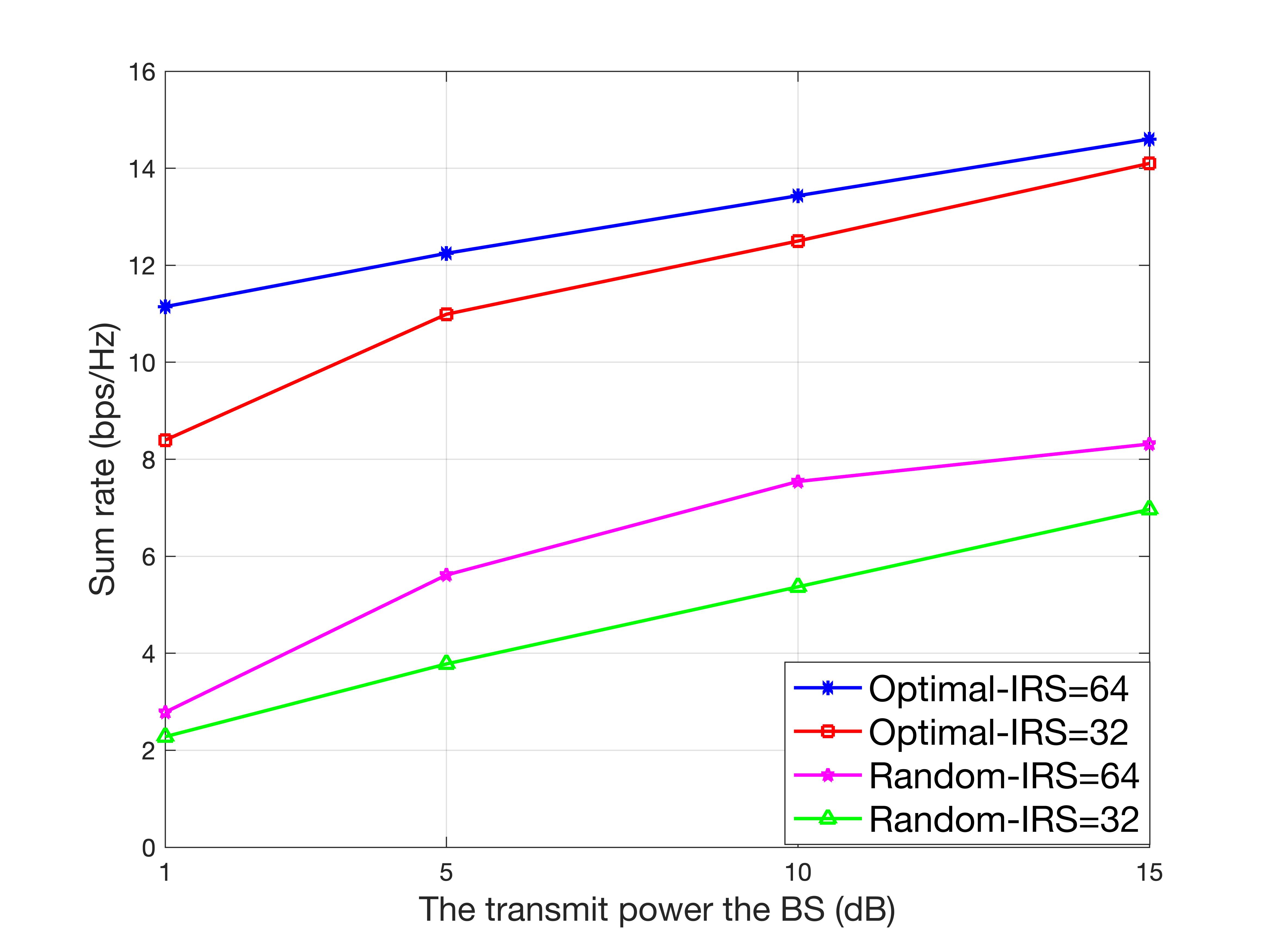}
\end{minipage}%
}%
\subfigure[IRS element number]{
\begin{minipage}[t]{0.5\linewidth}
\centering
\includegraphics[width=2in]{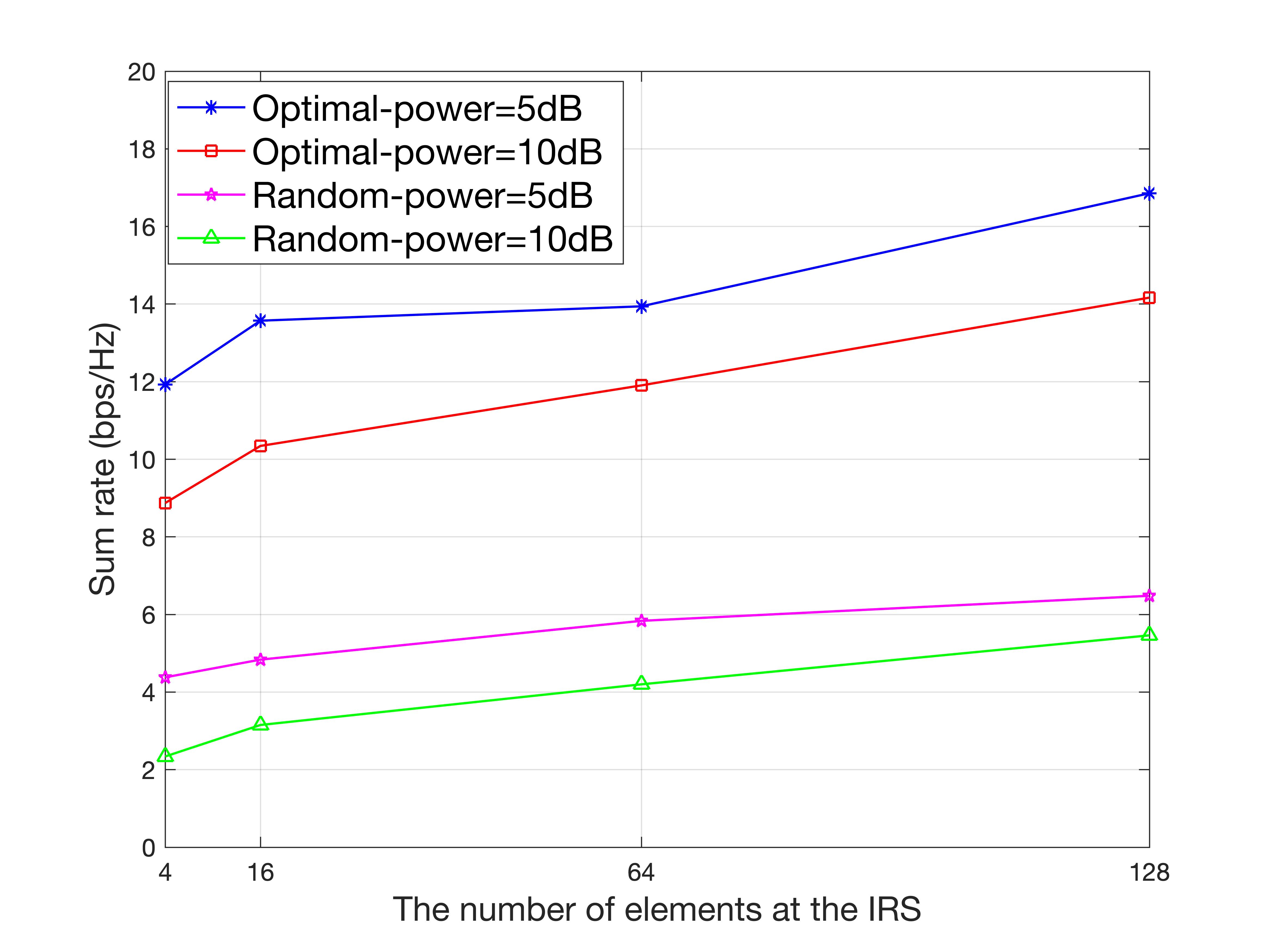}
\end{minipage}%
}%
\captionsetup{font={small}}
\caption{The performance of the proposed algorithm } \label{performance}
\vspace{-0.4cm}
\end{figure}

Fig. \ref{performance} shows the performance of the proposed algorithm. In these two subfigures, the X-axis represents the total transmit power at the BS and the number of the IRS elements, respectively. The Y-axis represents the sum rate of all the users for both two subfigures. It is obvious that the perforce is getting better with  the increasing of total transmit power and the number of IRS elements. It also shows that the beamforming and the phase shift optimized through DDPG outperform the random beamforming and phase shift.

\section{Conclusion}
In this letter, a sum rate maximization problem in an IRS assisted NOMA downlink network was investigated. A DDPG based algorithm was proposed to jointly optimize beamforming and phase shift. The proposed DDPG algorithm can not only achieve competitive performance but also adapt to the varying channel scenario, however, the conventional convex optimization is mainly suitable for the fixed channel scenario. More specifically, machine learning provides a new solution for wireless communication problems and also can be applied for more complicated scenarios, which will be a powerful tool for developing the next generation communication network.

\bibliographystyle{IEEEtran}
\bibliography{IEEEfull,EEref}

\begin{thebibliography}{10}
\providecommand{\url}[1]{#1}
\csname url@samestyle\endcsname
\providecommand{\newblock}{\relax}
\providecommand{\bibinfo}[2]{#2}
\providecommand{\BIBentrySTDinterwordspacing}{\spaceskip=0pt\relax}
\providecommand{\BIBentryALTinterwordstretchfactor}{4}
\providecommand{\BIBentryALTinterwordspacing}{\spaceskip=\fontdimen2\font plus
\BIBentryALTinterwordstretchfactor\fontdimen3\font minus
  \fontdimen4\font\relax}
\providecommand{\BIBforeignlanguage}[2]{{%
\expandafter\ifx\csname l@#1\endcsname\relax
\typeout{** WARNING: IEEEtran.bst: No hyphenation pattern has been}%
\typeout{** loaded for the language `#1'. Using the pattern for}%
\typeout{** the default language instead.}%
\else
\language=\csname l@#1\endcsname
\fi
#2}}
\providecommand{\BIBdecl}{\relax}
\BIBdecl

\bibitem{you2021towards}
X.~You, C.~Wang, J.~Huang, X.~Gao, Z.~Zhang, M.~Wang, Y.~Huang, C.~Zhang,
  Y.~Jiang, J.~Wang \emph{et~al.}, ``Towards {6G} wireless communication
  networks: Vision, enabling technologies, and new paradigm shifts,''
  \emph{Science China Information Sciences}, vol.~64, no.~1, pp. 1--74, 2021.

\bibitem{ding2015application}
Z.~Ding, F.~Adachi, and H.~V. Poor, ``The application of {MIMO} to
  non-orthogonal multiple access,'' \emph{IEEE Transactions on Wireless
  Communications}, vol.~15, no.~1, pp. 537--552, 2015.

\bibitem{saito2015performance}
K.~Saito, A.~Benjebbour, Y.~Kishiyama, Y.~Okumura, and T.~Nakamura,
  ``Performance and design of {SIC} receiver for downlink noma with open-loop
  {SU-MIMO},'' in \emph{2015 IEEE International Conference on Communication
  Workshop (ICCW)}.\hskip 1em plus 0.5em minus 0.4em\relax IEEE, 2015, pp.
  1161--1165.

\bibitem{8811733}
Q.~{Wu} and R.~{Zhang}, ``Intelligent reflecting surface enhanced wireless
  network via joint active and passive beamforming,'' \emph{IEEE Transactions
  on Wireless Communications}, vol.~18, no.~11, pp. 5394--5409, 2019.

\bibitem{huang2020reconfigurable}
C.~Huang, R.~Mo, and C.~Yuen, ``Reconfigurable intelligent surface assisted
  multiuser {MISO} systems exploiting deep reinforcement learning,'' \emph{IEEE
  Journal on Selected Areas in Communications}, vol.~38, no.~8, pp. 1839--1850,
  2020.

\bibitem{gao2020unsupervised}
J.~Gao, C.~Zhong, X.~Chen, H.~Lin, and Z.~Zhang, ``Unsupervised learning for
  passive beamforming,'' \emph{IEEE Communications Letters}, vol.~24, no.~5,
  pp. 1052--1056, 2020.

\bibitem{arulkumaran2017deep}
K.~Arulkumaran, M.~P. Deisenroth, M.~Brundage, and A.~A. Bharath, ``Deep
  reinforcement learning: A brief survey,'' \emph{IEEE Signal Processing
  Magazine}, vol.~34, no.~6, pp. 26--38, 2017.

\bibitem{zuo2020resource}
J.~Zuo, Y.~Liu, Z.~Qin, and N.~Al-Dhahir, ``Resource allocation in intelligent
  reflecting surface assisted {NOMA} systems,'' \emph{IEEE Transactions on
  Communications}, vol.~68, no.~11, pp. 7170--7183, 2020.

\bibitem{ding2020simple}
Z.~Ding and H.~V. Poor, ``A simple design of {IRS-NOMA} transmission,''
  \emph{IEEE Communications Letters}, vol.~24, no.~5, pp. 1119--1123, 2020.

\bibitem{franccois2018introduction}
V.~Fran{\c{c}}ois-Lavet, P.~Henderson, R.~Islam, M.~G. Bellemare, and
  J.~Pineau, ``An introduction to deep reinforcement learning,'' \emph{arXiv
  preprint arXiv:1811.12560}, 2018.

\end{thebibliography}
\vspace{0.5em}
\end{document}